\documentclass[
nofootinbib,
reprint,
 amsmath,amssymb,
 aps,
]{revtex4-2}

\usepackage{graphicx}% Include figure files
\usepackage{dcolumn}% Align table columns on decimal point
\usepackage{bm}% bold math
\def\p{\mbox{\boldmath$\displaystyle\boldsymbol{p}$}}
\def\k{\mbox{\boldmath$\displaystyle\boldsymbol{k}$}}

\def\bv{\mbox{\boldmath$\displaystyle\boldsymbol{\varphi}$}}

\def\0{\mbox{\boldmath$\displaystyle\boldsymbol{0}$}}
\def\s{\mbox{\boldmath$\displaystyle\boldsymbol{\sigma}$}}

\def\x{\mbox{\boldmath$\displaystyle\boldsymbol{x}$}}
\def\y{\mbox{\boldmath$\displaystyle\boldsymbol{y}$}}

\newcommand{\dual}[1]{\overset{\:{}^{^{{{\neg}}}}}{\smash[t]{#1}}} %Elko dual 

\newcommand{\dualn}[1]{\overset{\:{}^{^{^{{{\neg}}}}}}{\smash[t]{#1}}} %Elko dual 

\begin{document}

\preprint{APS/123-QED}

\title{\boldmath Mass dimension one fermions in FLRW space-time}% Force line breaks with \\

\author{Cheng-Yang Lee}
\email{cylee@scu.edu.cn} 
\affiliation{Center for theoretical physics, College of Physics,\\ Sichuan University, Chengdu, 610064, China } 
 
\author{Haomin Rao}
\email{rhm137@mail.ustc.edu.cn}
\affiliation{University of Chinese Academy of Sciences, 100190 Beijing, China}
\affiliation{School of Fundamental Physics and Mathematical Sciences,
Hangzhou Institute for Advanced Study, UCAS, Hangzhou 310024, China}

\author{Wenqi Yu}
\email{wyuaz@connect.ust.hk}
\affiliation{The Hong Kong University of Science and Technology, Clear Water Bay, Kowloon, Hong Kong, P.R.China} 
\affiliation{The HKUST Jockey Club Institute for Advanced Study,
The Hong Kong University of Science and Technology,
Clear Water Bay, Kowloon, Hong Kong, P.R. China}
\author{Siyi Zhou}
\email{siyi@cqu.edu.cn} 
\affiliation{Department of Physics and Chongqing Key Laboratory for Strongly Coupled Physics,
Chongqing University, Chongqing 401331, China} 

\date{\today}% It is always \today, today,

\begin{abstract}
Elko is a massive spin-half field of mass dimension one. Elko differs from the Dirac and Majorana fermions because it furnishes the irreducible representation of the extended Poincar\'{e} group with a two-fold Wigner degeneracy where the particle and anti-particle states both have four degrees of freedom. Elko has a renormalizable quartic self interaction which makes it a candidate for self-interacting dark matter. We study Elko in the spatially flat FLRW space-time and find exact solutions in the de Sitter space. By choosing the appropriate solutions and phases, the fields satisfy the canonical anti-commutation relations and have the correct time evolutions in the flat space limit. 
\end{abstract}

\maketitle

%\tableofcontents

\section{Introduction}
The energy scale of inflation can be as high as $10^{13}$ GeV. It is therefore natural to use inflation to probe high energy physics which is beyond the reach of the current Large Hadron Collider (LHC). This program is called the cosmological collider physics\cite{Chen_2010qsfi,Chen:2010xka,Komatsu:2010hc,Barnaby:2011vw,Chen:2011zf,Chen:2012ge,Baumann_2012,Noumi_2013,Arkani-Hamed:2015bza,Lee_2016,Arkani-Hamed:2017fdk,Arkani-Hamed:2018kmz,meerburg2019primordial,Baumann:2019oyu,Baumann:2020dch} and it has often been used to probe physics beyond the Standard Model (SM)\footnote{See~\cite{Werth:2024aui,McCulloch:2024hiz,Craig:2024qgy,Chen:2023iix} for recent developments of this program.}. This framework can be used to detect various types of particles, including massive scalars~\cite{Chen_2010qsfi,Chen_2010l,Chen_2012}, vectors~\cite{Liu:2015tza,Wang_2020},  graviton~\cite{Tong_2022}, higher spin particles~\cite{Arkani-Hamed:2015bza} and massive fermions~\cite{Chen_2018}. It can also be used to detect massless particles as well as the SM particles~\cite{Hook_2020}. 

While the SM has been successful in explaining most of the phenomenologies at the LHC, there are strong motivations to study physics beyond the SM. 
%For example, the hierarchy problem has led physicists to propose the little Higgs model and search for supersymmetric particles. 
In particular, according to the $\Lambda$CDM model, the SM particles can only account for 5\% of the total energy and mass contents in the observable universe with the remaining 25\% and 70\% appear in the form of dark matter and dark energy. The task of searching and finding the correct theory of dark matter are among the most important problems in fundamental physics.

%A large proportion of the mass cannot be explained by the SM. These are known as the dark matter. There are wide interest in the beyond standard model community.

In 2004, Ahluwalia and Grumiller proposed the theory of Elko and mass dimension one fermionic fields of spin-half~\cite{Ahluwalia:2004ab,Ahluwalia:2004sz}.\footnote{Elko is the German acronym Eigenspinoren des Ladungskonjugatioperators. In English, it means the eigenspinors of the charge-conjugation operator. By definition, Elko is a four-component spinor in the $\left(\frac{1}{2},0\right)\oplus\left(0,\frac{1}{2}\right)$ representation. But for expedience, whenever no confusions arise, we will simply use the term Elko to mean mass dimension one fields.} In the coming decades, much works have been devoted to refine the construct~\cite{Ahluwalia:2008xi,Ahluwalia:2009rh,Ahluwalia:2010zn,Lee:2012td,Lee:2014opa,Lee:2015sqj,Ahluwalia:2016rwl,Lee:2019fni,Ahluwalia:2019etz,Ahluwalia:2022ttu}, finally yielding a local theory that respects Lorentz symmetry~\cite{Ahluwalia:2022yvk,Ahluwalia:2023slc}. These fermionic fields are physically distinct from their Dirac counterparts. From a group representation perspective, they furnish the irreducible representation of the extended Poincar\'{e} group with a two-fold Wigner degeneracy. What this means is that the particle and anti-particle states each have four degrees of freedom, labelled by the spin-projection $\sigma=\pm\frac{1}{2}$ and an extra degeneracy index $n=\pm1$. Elko has mass dimension one because its kinematics is described by the spinorial Klein-Gordon and not the Dirac equation. Consequently, these fermions have renormalizable quartic self-interaction, a desirable property for dark matter. 

There were attempts in building phenomenological models~\cite{Dias:2010aa,Alves:2014kta,Alves:2014qua,Agarwal:2014oaa,Alves:2017joy,Moura:2021rmf}, but they were premature. Firstly, the free fields had the long-standing problem of rotational symmetry breaking which perpetuates to the interaction, rendering these models physically undesirable (there are stringent experimental limits on rotational symmetry breaking). This problem was only resolved recently~\cite{Ahluwalia:2022yvk}. Secondly, the Elko interactions are non-Hermitian. More precisely, they are pseudo-Hermitian. Works that have taken pseudo-Hermiticity into account can be found in~\cite{Ahluwalia:2022ttu,deGracia:2023yit,Ahluwalia:2023slc} though more efforts are needed to establish self-consistency. 
%Therefore, one must develop the appropriate formalism to compute physical observables. 

In this paper, we initiate a systematic study of Elko in curved space-time. The ultimate objective is to derive observables for Elko in the cosmological collider framework. Gravitational dynamics originating from the interplay of Klein-Gordon kinematics and spin-connections have been extensively studied in cosmology~\cite{Boehmer:2006qq,Boehmer:2007dh,Boehmer:2008rz,Boehmer:2008ah,Boehmer:2009aw,Boehmer:2010ma,Basak:2012sn,Sadjadi:2012xyd,HoffdaSilva:2014tth,Pereira:2014wta,S:2014woy,Pereira:2014pqa,Chang:2015ufa,Pereira:2016emd,Pereira:2016eez,Pereira:2017efk} and higher-dimensional space-time~\cite{Liu:2011nb,Zhou:2017bbj,Sorkhi:2018jhy,Zhou:2020ucc}. But to the best of our knowledge, these works have not directly confronted the issue of canonical quantization. What we will accomplish here, is to study Elko in the spatially flat FLRW background. We will solve the equations of motion in the de Sitter (dS) space and perform canonical quantization. In the process, the main difficulty is that we have to solve a system of coupled differential equations. This difficulty is circumvented by making the observation that Elko has a dual-helicity structure. That is, the right- and left-handed components of Elko have opposite helicity. Utilizing the dual-helicity structure, the differential equations decouple and admit exact solutions that are amenable to consistent quantization in the dS background with physically well-defined time evolutions in the flat space limit. 
%As Elko has Klein-Gordon kinematics, considerable attentions have been devoted to study Elko as an inflaton.

This paper is organized as follows. In sec.~\ref{elko}, we review the theory of Elko and mass dimension one fermions in Minkowski space-time. In sec.~\ref{elkofrw}, we formulate the theory in the FLRW background following~\cite{HoffdaSilva:2014tth} and make the observation that in curved space-time, the spinorial Klein-Gordon equation cannot be obtained by simply replacing the partial derivative by the covariant derivative. The reason being $(\gamma^{\mu}\nabla_{\mu})^{2}$ is not equal to $g^{\mu\nu}\nabla_{\mu}\nabla_{\nu}$. The relationship between these two operators are given by the Lichnerowicz formula~\cite{Lichnerowicz} (see app.~\ref{Lichnerowicz} for details). In sec.~\ref{quantizationindS}, we solve the equations of motion for Elko in dS space and canonically quantize them. By choosing the appropriate solutions and phases, we show that Elko satisfies the canonical anti-commutation relations and have the correct time evolutions in the flat space limit. Conclusion and outlook are given in sec.~\ref{conc}.  

%Instead, the correct equation of motion requires us to add a factor of $R/4$ to the mass square term where $R$ is the Ricci scalar
\section{Elko in flat space-time}\label{elko}
%------------------------------------------------------------------

We review the theory of Elko and mass dimension one fermionic fields in the Minkowski space-time. For more details, see~\cite{Ahluwalia:2022yvk}. Elko are eigenspinors of the  charge conjugation operator in the $\left(\frac{1}{2},0\right)\oplus\left(0,\frac{1}{2}\right)$ representation. The mass dimension one fermionic fields are quantum fields constructed using Elko as expansion coefficients.

We work in the representation where the $\gamma^{\mu}$ matrices are given by~\cite{Weinberg:1995mt}
\begin{equation}
\gamma^{0}=-i\left(\begin{matrix}
    \mathbb{O}_{2} & \mathbb{I}_{2} \\
    \mathbb{I}_{2} & \mathbb{O}_{2}
\end{matrix}\right),\quad
\gamma^{i}=-i\left(\begin{matrix}
    \mathbb{O}_{2} &-\sigma^{i} \\
    \sigma^{i} & \mathbb{O}_{2}
\end{matrix}\right),
\end{equation}
and $\gamma^{5}=i\gamma^{0}\gamma^{1}\gamma^{2}\gamma^{3}$. We take the Minkowski metric to be $(-,+,+,+)$, so the anti-commutators of $\gamma^{a}$ are
\begin{equation}
    \left\{\gamma^{a},\gamma^{b}\right\}=2\eta^{ab}\mathbb{I}_{4},
\end{equation}
where $\eta^{00}=-1$ and $\eta^{ij}=\delta^{ij}$. We use the Latin ($a,b,\cdots$) and Greek alphabets ($\alpha,\beta,\cdots)$ to label objects in Minkowski and in curved space-time respectively. The Lorentz generators are
\begin{equation}
\Sigma^{ab}=-\frac{i}{4}\left[\gamma^{a},\gamma^{b}\right].\label{eq:sigma_ab}
\end{equation}
Its respective rotation and boost generators are
\begin{equation}
    \pmb{\mathcal{J}}=(\Sigma^{23},\Sigma^{31},\Sigma^{12})=\frac{1}{2}\left(\begin{matrix}
        \boldsymbol{\sigma} & \pmb{\mathbb{O}}_{2} \\
        \pmb{\mathbb{O}}_{2} & \boldsymbol{\sigma}
    \end{matrix}\right),
\end{equation}
and
\begin{equation}
    \pmb{\mathcal{K}}=(\Sigma^{01},\Sigma^{02},\Sigma^{03})=
    \frac{i}{2}\left(\begin{matrix}
        \boldsymbol{\sigma} & \pmb{\mathbb{O}}_{2} \\
        \pmb{\mathbb{O}}_{2} &-\boldsymbol{\sigma}
    \end{matrix}\right),
\end{equation}
where $\s=(\sigma^{1},\sigma^{2},\sigma^{3})$ are the Pauli matrices. The boost and rotation transformations of spinors in the $\left(\frac{1}{2},0\right)\oplus\left(0,\frac{1}{2}\right)$ representation are given by $\exp(-i\pmb{\mathcal{K}}\cdot\bv)$ and $\exp(i\pmb{\mathcal{J}\cdot\theta})$ where $\bv=\varphi\hat{\k}$, $\boldsymbol{\theta}=\theta\boldsymbol{\hat{n}}$ and $\cosh\varphi=\omega_{k}/m$, $\sinh\varphi=|\k|/m$ with $\omega_{k}=\sqrt{|\k|^{2}+m^{2}}$. Here, we adopt the convention where the top and bottom component of the spinors to be right-handed and left-handed respectively.

The mass dimension one fermionic field $\lambda$ and its dual $\dual{\lambda}$ are given by
\begin{equation}
\begin{aligned}
    &\lambda(x)=\int\frac{d^{3}k}{(2\pi)^{3}}\frac{1}{\sqrt{2m\omega_{k}}}\sum^{4}_{\tau=1}
    \left[e^{ik\cdot x}\xi_{\tau}(\k)a_{\tau}(\k)+\right.\\
    &\left. e^{-ik\cdot x}\chi_{\tau}(\k)b^{\dag}_{\tau}(\k)\right], 
\end{aligned}
    \label{eq:lambda}
\end{equation}
and
\begin{equation}
\begin{aligned}
    &\dual{\lambda}(x)=\int\frac{d^{3}k}{(2\pi)^{3}}\frac{1}{\sqrt{2m\omega_{k}}}\sum^{4}_{\tau=1}
    \left[e^{-ik\cdot x}\dual{\xi}_{\tau}(\k)a^{\dag}_{\tau}(\k)+\right.\\
    &\left.e^{ik\cdot x}\dual{\chi}_{\tau}(\k)b_{\tau}(\k)\right].
\end{aligned}    
\end{equation}
As shown in~\cite{Ahluwalia:2022yvk}, rotational symmetry and locality require the particle and anti-particle states to each have four degrees of freedom $\tau=1,\cdots,4$. In~(\ref{eq:lambda}), the spinors $\xi$ and $\chi$ are Elko, having eigenvalues $+1$ and $-1$ with respect to the charge-conjugation operator
\begin{equation}
    \mathcal{C}\xi_{\tau}(\k)=+\xi_{\tau}(\k),\quad
    \mathcal{C}\chi_{\tau}(\k)=-\chi_{\tau}(\k)
\end{equation}
where
\begin{equation}
	\mathcal{C}=\left(\begin{array}{cc}
	\mathbb{O}_{2} & i\Theta \\
	-i\Theta & \mathbb{O}_{2}
	\end{array}\right) K,
\end{equation}
with
\begin{equation}
    \Theta=\left(\begin{matrix}
        0 & -1 \\
        1 & 0
    \end{matrix}\right)
\end{equation}
being the Wigner time-reversal matrix which satisfies $\Theta\s\Theta^{-1}=-\s^{*}$and $K$ the complex conjugation operator acting to its right $Kf=f^{*}$. The solutions of Elko are given by
\begin{equation}
    \begin{aligned}
\xi_{1}(\k)&=\left[\begin{array}{r}
	+i \phi_{-}(\k) \\
	\phi_{+}(\k)
	\end{array}\right], \quad
 \xi_{2}(\k)=\left[\begin{array}{r}
	-i\phi_{+}(\k) \\
	 \phi_{-}(\k)
	\end{array}\right],\label{eq:xi}\\
  \xi_{3}(\k)&=\left[\begin{array}{r}
	\phi_{+}(\k) \\
	-i \phi_{-}(\k)
	\end{array}\right], \quad 
 \xi_{4}(\k)=\left[\begin{array}{r}
	\phi_{-}(\k) \\
	+i\phi_{+}(\k)
	\end{array}\right], 
 \end{aligned}
\end{equation}

 and 
\begin{equation}
    \begin{aligned}
\chi_{1}(\k)&=\left[\begin{array}{r}
	+i \phi_{+}(\k) \\
	\phi_{-}(\k)
	\end{array}\right], \quad
 \chi_{2}(\k)=\left[\begin{array}{r}
	i\phi_{-}(\k) \\
	 -\phi_{+}(\k)
	\end{array}\right],\\
  \chi_{3}(\k)&=\left[\begin{array}{r}
	\phi_{-}(\k) \\
	-i\phi_{+}(\k)
	\end{array}\right], \quad 
 \chi_{4}(\k)=\left[\begin{array}{r}
	-\phi_{+}(\k) \\
	-i \phi_{-}(\k)
	\end{array}\right],\label{eq:zeta}
 \end{aligned}
\end{equation}
where $\phi_{\pm}$ are the Weyl spinors. Here, we work in the helicity basis
\footnote{we could choose other basis such as polarization basis, but in that cases the direction of the spinors will change under a boost transformation, so it is inconvenient to compute.}
so that $\phi_{\pm}(\k)$ are eigenspinors of $\s\cdot\k$ with eigenvalue $\pm|\k|$. In the rest frame, they are given by
\begin{equation}
\begin{aligned}
    &\phi_{+}(\boldsymbol{\epsilon})=\sqrt{\frac{m}{2}}\left(\begin{matrix}
        c_{\theta/2}  e^{-\frac{i}{2}\phi}\\
        s_{\theta/2}  e^{\frac{i}{2}\phi}
    \end{matrix}\right),\\
    &\phi_{-}(\boldsymbol{\epsilon})=\sqrt{\frac{m}{2}}\left(\begin{matrix}
        -s_{\theta/2} e^{-\frac{i}{2}\phi}\\
        c_{\theta/2}  e^{\frac{i}{2}\phi} \label{eq:weyl_spinors}
    \end{matrix}\right),
\end{aligned}
\end{equation}
where $s_{\theta}=\sin\theta$, $c_{\theta}=\cos\theta$ and $\boldsymbol{\epsilon}=(s_{\theta}c_{\phi},s_{\theta}s_{\phi},c_{\theta})$ specifies the direction of the spin-projection in the spherical coordinate such that $(\s\cdot\boldsymbol{\epsilon})\phi_{\pm}(\boldsymbol{\epsilon})=\pm\phi_{\pm}(\boldsymbol{\epsilon})$. In the helicity basis, the direction of boost is $\boldsymbol{\epsilon}$. As evident from~(\ref{eq:xi}-\ref{eq:zeta}), the left- and right-handed components of Elko have opposite helicity eigenvalues with respect to $\s\cdot\k$. This feature is known as the \textit{dual-helicity} structure and it will play an important role in sec.~\ref{quantizationindS} when we solve the equations of motion for Elko in the dS space. The Elko duals are given by
\begin{equation}
  \begin{aligned}
    \dual{\xi}_{1}(\k)&=-i\xi^{\dag}_{2}(\k)\beta,\quad &\dual{\xi}_{2}(\k)=+i\xi^{\dag}_{1}(\k)\beta,\label{eq:dual_xi1}\\
    \dual{\xi}_{3}(\k)&=+i\xi^{\dag}_{4}(\k)\beta,\quad &\dual{\xi}_{4}(\k)=-i\xi^{\dag}_{3}(\k)\beta,
\end{aligned}  
\end{equation}
and
\begin{equation}
  \begin{aligned}
    \dual{\chi}_{1}(\k)&=+i\chi^{\dag}_{2}(\k)\beta,\quad &\dual{\chi}_{2}(\k)=-i\chi^{\dag}_{1}(\k)\beta,\\
    \dual{\chi}_{3}(\k)&=-i\chi^{\dag}_{4}(\k)\beta,\quad &\dual{\chi}_{4}(\k)=+i\chi^{\dag}_{3}(\k)\beta,\label{eq:dual_zetz4}
\end{aligned}  
\end{equation}
where $\beta\equiv i\gamma^{0}$. Direct evaluations show that Elko are orthonormal
\begin{equation}
	\dual{\xi}(\k, \tau) \xi(\k, \tau')=-\dual{\chi}(\k,\tau) \chi(\k,\tau')=m  \delta_{\tau\tau'},
\end{equation}
and the spin-sums are
\begin{equation}
	\begin{aligned}
	& \sum^{4}_{\tau=1} \xi_{\tau}(\k) \dual{\xi}_{\tau}(\k)=+ m \mathbb{I}_4 \\
	& \sum^{4}_{\tau=1} \chi_{\tau}(\k) \dual{\chi}_{\tau}(\k)=-m  \mathbb{I}_4.
	\end{aligned}
\end{equation}

Using the spin-sums, we find $\{\lambda(t,\x),\dual{\lambda}(t,\y)\}=\mathbb{O}_{4}$. The free propagator obtained from the two-point time-ordered product is fermionic and is given by
\begin{equation}
	i\langle 0|T\lambda\left(x'\right) \dual{\lambda}(x)|0\rangle=\int \frac{d^4 k}{(2 \pi)^4} e^{i k\cdot\left(x^{\prime}-x\right)} \frac{\mathbb{I}_{4}}{k^{2}+m^2-i \epsilon}.
\end{equation}
Therefore, the free Lagrangian density takes the form
\begin{equation}
	\mathcal{L}_0=-\left[(\partial^a \dual{\lambda})(\partial_a \lambda)+m^2 \dual{\lambda}\lambda\right].
\end{equation}
From the conjugate momentum $\pi=\partial_{t}\dual{\lambda}$, we readily verify
%\begin{equation}
%\left\{\lambda(t,\x),\dual{\lambda}(t,\y)\right\}=\mathbb{O}_{4},
%\end{equation}
the canonical anti-commutation relations
\begin{equation}
\begin{aligned}
    \left\{\lambda(t,\x),\lambda(t,\x')\right\}&=\mathbb{O}_{4},\\
    \left\{\pi(t,\x),\pi(t,\x')\right\}&=\mathbb{O}_{4},\\
	\left\{\lambda(t, \x), \pi(t, \x')\right\}&=i\delta^{3}(\x-\x')\mathbb{I}_{4}.
 \end{aligned}
\end{equation}
The free Hamiltonian evaluates to  
\begin{align}
	H_0&=\int\frac{d^{3}k}{(2\pi)^{3}}\left[(\partial_{t}\dual{\lambda})(\partial_{t}\lambda)+(\nabla \dual{\lambda})\cdot(\nabla \lambda)+m^2\dual{\lambda}\lambda\right] \nonumber\\
    &=\int\frac{d^{3}k}{(2\pi)^{3}}\omega_{k}\sum^{4}_{\tau=1}\left[a^{\dagger}_{\tau}(\k)a_{\tau}(\k)-b_{\tau}(\k)b^{\dagger}_{\tau}(\k)\right].
\end{align}
It is positive-definite (with negative vacuum energy) provided that the annihilation and creation operators satisfy fermionic statistics
\begin{equation}
    \left\{a_{\tau}(\p),a^{\dag}_{\sigma}(\p')\right\}
    =\left\{b_{\tau}(\p),b^{\dag}_{\sigma}(\p')\right\}=(2\pi)^{3}\delta_{\tau\sigma}\delta^{3}(\p-\p').\label{eq:car}
\end{equation}

Since $\dual{\lambda}$ and $\lambda$ are of mass dimension one, Elko has a renormalizable quartic self interaction
$g(\dual{\lambda}\lambda)^2$ which makes it a candidate for self-interacting dark matter. Elko can also interact with the Higgs boson. Apart from the Yukawa interaction $g\dual{\lambda}\lambda\phi$, we also have renormalizable interactions of the form $g'\dual{\lambda}\lambda\phi^{2}$. Prior to study Elko phenomenology, we have to deal with the issue of non-Hermitian interactions. We note, while the free Hamiltonian is Hermitian, the individual terms which contribute to $H_{0}$ are non-Hermitian. If we study interactions such as the ones presented here, then Hermiticity is lost thus making the time evolutions non-unitary. Therefore, a new formalism to compute transition probabilities and observables has to be developed. Recent work establishing pseudo Hermiticity of Elko provides a promising approach to resolve this problem~\cite{Ahluwalia:2023slc}. This task is important but beyond the scope of the present work. The primary focus here is to study Elko in the spatially flat FLRW space-time and show that the free fields can be consistently quantized. Interactions in the cosmological collider framework will be studied in the future.

%In the absence of interactions, we do not have to deal with non-Hermiticity. 
%show accomplish here is to show that in curved space-time, 

Before we proceed to study Elko in curved space-time, there is an important issue concerning the solutions ~(\ref{eq:xi}-\ref{eq:zeta}) that requires further elaborations. Readers who are knowledgeable in this subject would have noticed that we have not presented Elko in its traditional form. In previous works, if left-handed component of Elko is $\phi_{\pm}$, then its right-handed component takes the form $\vartheta\Theta\phi^{*}_{\pm}$ where $\vartheta$ is an imaginary phase taken to be $+i$ or $-i$. What the traditional construct implies is that the left- and right-handed components are not independent variables. If we confine the discussions within the Lorentz group, then nothing goes wrong. That is, given a left-handed Weyl spinor $\phi_{\pm}$, then $\Theta\phi^{*}_{\pm}$ transforms as a right-handed spinor under boost and rotation. On the other hand, if we include space-time translation, this argument fails. Because under space-time translation, the left- and right-handed spinors transform differently $\phi_{\pm}\rightarrow e^{-ik\cdot x}\phi_{\pm}$, $\Theta\phi^{*}_{\pm}\rightarrow e^{ik\cdot x}\Theta\phi^{*}_{\pm}$. In light of this observation, we believe that the correct approach is to simply treat Elko as a set of four-component spinors with dual-helicity structure where its left- and right-handed component are independent variables. Since it has already been verified that Elko furnishes the irreducible representation of the extended Poincar\'{e} group with a two-fold Wigner degeneracy~\cite{Ahluwalia:2022yvk,Ahluwalia:2023slc}, the traditional construct starting from the two-component Weyl spinors is no longer necessary. In sec.~\ref{quantizationindS}, when solving the equations of motion for Elko in the spatially flat FLRW space-time, this issue appears once more. As we will demonstrate, if the left- and right-handed components of Elko are related according to the traditional approach, then the equations of motion only admit trivial solution.

\section{Elko in curved space-time}\label{elkofrw}

In the Minkowski space-time, the equation of motion for Elko is the spinorial Klein-Gordon equation. In curved space-time, Elko satisfies
\begin{align}
    \left(\gamma^{\mu}\nabla_{\mu}\gamma^{\nu}\nabla_{\nu}-m^{2}\right)\lambda=0. \label{eq:eom}
\end{align}
Therefore, the action is
\begin{equation}
S=-\int d^4 x \sqrt{ -g}\left[\nabla_\mu \dual{\lambda} \mathbf{\gamma^\mu \gamma^\nu} \nabla_\nu \lambda+m^{2}\dual{\lambda} \lambda\right].
\end{equation}
The motivation for writing the equation of motion using $\gamma^{\mu}\nabla_{\mu}$ instead of the Laplacian $g^{\mu\nu}\nabla_{\mu}\nabla_{\nu}$ will be explained below. The Dirac matrices satisfy
\begin{equation}
    \left\{\gamma^{\mu},\gamma^{\nu}\right\}=2g^{\mu\nu}\mathbb{I}_{4},
\end{equation}
where $g^{\mu\nu}$ is the metric. We take the local and global coordinates to be labelled by the Latin ($a,b,\cdots$) and Greek ($\mu,\nu,\cdots$) alphabets respectively. The Dirac matrices in the two coordinates are related by
\begin{equation}
\begin{aligned}
    \gamma^\mu={e_a}^\mu \gamma^a, \\
\gamma_\mu={e^a}_\mu \gamma_a,
\end{aligned}
\end{equation}

where ${e^{a}}_{\mu}$ is the tetrad
	\begin{equation}
\begin{aligned}
	g_{\mu\nu}&={e^{a}}_{\mu}{e^{b}}_{\nu}\eta_{ab},\\
    \eta_{ab}&={e_{a}}^{\mu}{e_{b}}^{\nu}g_{\mu\nu}.
	\end{aligned}
\end{equation}
The covariant derivatives acting on the vectors and spinors are \cite{yepez2011einsteins}
\begin{equation}
\begin{aligned}
%\nabla_\mu X^\nu & =\partial_\mu X^\nu+\Gamma_{\mu \lambda}^\nu X^\lambda, \\
%\nabla_\mu X_\nu & =\partial_\mu X_\nu-\Gamma_{\mu \nu}^\lambda X_\lambda, \\
%\nabla_\mu X^a & =\partial_\mu X^a+\omega_\mu{ }^a{ }_b X^b, \\ 
%\nabla_\mu X_a & =\partial_\mu X_a-\omega_\mu{ }^b{ }_a X_b, \\
\nabla_\mu\dual{\lambda}&=\partial_{\mu}\dual{\lambda}-\dual{\lambda}\Gamma_{\mu}, \label{eq:cov1}\\    
\nabla_\mu \lambda&=\partial_\mu \lambda+\Gamma_\mu \lambda,
\end{aligned}
\end{equation}
where $\Gamma^{\nu}_{\mu\lambda}$ is the Christoffel symbol and
\begin{equation}
\Gamma_\mu =\frac{i}{2}{\omega_{\mu}}^{ab}\Sigma_{ab}.
\end{equation}
The spin connection ${\omega_{\mu}}^{ab}$ and the Christoffel symbol are related by
\begin{equation}
{\omega_{\mu}}^{ab}={e^{a}}_{\nu}\left(\partial_{\mu}e^{b\nu}+e^{b\sigma}\Gamma^{\nu}_{\mu\sigma}\right).
\end{equation}
The metric and tetrad satisfy
\begin{equation}
    \nabla_{\sigma}g_{\mu\nu}=0,\quad\nabla_{\sigma}{e^{a}}_{\nu}=0.\label{eq:dg}
\end{equation}

The main objective of this paper is to find exact solutions for Elko in the dS space and show that it can be consistently quantized. These tasks will be accomplished in the next section. To solve~(\ref{eq:eom}), we first rewrite it in terms of partial derivatives and make the following observation.
In curved space-time, the square of the Dirac operator $(\gamma^{\mu}\nabla_{\mu})^{2}$ is not equal to the Laplacian $g^{\mu\nu}\nabla_{\mu}\nabla_{\nu}$. Using the Lichnerowicz formula (see app.~\ref{Lichnerowicz} for details)~\cite{Lichnerowicz},
\begin{equation}
	    \gamma^{\mu}\nabla_{\mu}\gamma^{\nu}\nabla_{\nu}\lambda=\left(g^{\mu\nu}\nabla_{\mu}\nabla_{\nu}-\frac{1}{4}R\right)\lambda,
\end{equation}
where $R$ is the Ricci scalar, we obtain
\begin{align}
    \left[g^{\mu\nu}\nabla_{\mu}\nabla_{\nu}-\left(m^{2}+\frac{1}{4}R\right)\right]\lambda=0.\label{eq:Elko1}
\end{align}
Using~(\ref{eq:dg}) and~(\ref{eq:cov1}), the Laplacian is given by
\begin{align}
    g^{\mu\nu}\nabla_{\mu}\nabla_{\nu}\lambda&=\nabla_{\mu}\nabla^{\mu}\lambda \nonumber\\
    &=\frac{1}{\sqrt{-g}}\partial_{\mu}\left[\sqrt{-g}g^{\mu\nu}\nabla_{\nu}\lambda\right]+g^{\mu\nu}\Gamma_{\mu}\nabla_{\nu}\lambda.
\end{align}
The equation of motion becomes~\cite{HoffdaSilva:2014tth}
\begin{equation}
    \begin{aligned}
    &\partial_{\mu}\left[\sqrt{-g}g^{\mu\nu}(\partial_{\nu}\lambda+\Gamma_{\nu}\lambda)\right]+\\
    &\sqrt{-g}\left[g^{\mu\nu}\Gamma_{\mu}(\partial_{\nu}\lambda+\Gamma_{\nu}\lambda)-\left(m^{2}+\frac{1}{4}R\right)\lambda\right]=0.
    \label{eq:Elko3}
\end{aligned}
\end{equation}
Repeat the above calculations for $\dual{\lambda}$, we obtain
\begin{equation}
    \begin{aligned}
    &\partial_{\mu}\left[\sqrt{-g}g^{\mu\nu}(\partial_{\nu}\dual{\lambda}-\dual{\lambda}\Gamma_{\nu})\right]-\\
    &\sqrt{-g}\left[g^{\mu\nu}\Gamma_{\mu}(\partial_{\nu}\dual{\lambda}-\dual{\lambda}\Gamma_{\nu})-\left(m^{2}+\frac{1}{4}R\right)\dual{\lambda}\right]=0.
    \label{eq:Elko4}
\end{aligned}
\end{equation}
  
%Repeat the same calculations for $\dual{\lambda}$, we obtain
%\begin{equation}
%    \partial_{\mu}\left[\sqrt{-g}g^{\mu\nu}(\partial_{\nu}\dual{\lambda}-\dual{\lambda}\Gamma_{\nu})\right]-\sqrt{-g}\left[g^{\mu\nu}(\partial_{\nu}\dual{\lambda}-\dual{\lambda}\Gamma_{\nu})\Gamma_{\mu}+\left(m^{2}+\frac{1}{4}R\right)\dual{\lambda}\right]=0.\label{eq:Elko4}
%\end{equation}

\section{Quantization in the de Sitter space}\label{quantizationindS}

In this section, we study Elko in the spatially flat FLRW space-time where the metric is
\begin{equation}
	d s^2=-dt^2+a^2(t)\left(d x^2+d y^2+d z^2\right).\label{eq:flrw}
\end{equation}
%After obtaining the equations of motion associated with~(\ref{eq:flrw}), we solve them in the dS space and perform canonical quantization. %We show that the exact solutions, which are given by the Whittaker functions yields a local quantum field theory satisfying the canonical anti-commutation relations and exhibit the correct behavior in the flat space limit.
The spin connections for~(\ref{eq:flrw}) are
\begin{equation}
\Gamma_0=\mathbb{O}_{4},\quad \Gamma_i=\frac{1}{2}\dot{a}\gamma_0 \gamma_i. \label{eq:spin_c}
\end{equation}
Substituting~(\ref{eq:spin_c}) into~(\ref{eq:Elko3}-\ref{eq:Elko4}), we obtain~\cite{HoffdaSilva:2014tth}
\begin{equation}
   \begin{aligned}
	&\ddot{\lambda}+3\left(\frac{\dot{a}}{a}\right) \dot{\lambda}-\frac{1}{a^2}\partial^{i}\partial_{i} \lambda-\frac{3}{4}\left(\frac{\dot{a}}{a}\right)^2 \lambda+\\
 &\left(m^{2}+\frac{1}{4}R\right) \lambda+\frac{\dot{a}}{a^2} \gamma^0 \gamma^i\left(\partial_i \lambda\right)=0,\label{eq:eom1}
\end{aligned} 
\end{equation}
and
 \begin{equation}
   \begin{aligned}
    &\ddot{\dualn{\lambda}}+3\left(\frac{\dot{a}}{a}\right) \dot{\dualn{\lambda}}-\frac{1}{a^2} \partial^{i}\partial_{i} {\dualn{\lambda}}-\frac{3}{4}\left(\frac{\dot{a}}{a}\right)^2 {\dual{\lambda}}+\\
    &\left(m^{2}+\frac{1}{4}R\right){\dual{\lambda}}-\frac{\dot{a}}{a^2}\left(\partial_i {\dual{\lambda}}\right) \gamma^{0}\gamma^{i}=0 .\label{eq:eom2}
\end{aligned} 
\end{equation}

In curved space-time, we expand the fields as
\begin{equation}
    \begin{aligned}
    &\lambda(t,\x)=\int\frac{d^{3}k}{(2\pi)^{3}}\frac{1}{\sqrt{2mka^{3}(t)}}\sum^{4}_{\tau=1}
    \left[e^{i\boldsymbol{k\cdot x}}\xi_{\tau}(\k,t)a_{\tau}(\k)\right.\\
    &\left.+e^{-i\boldsymbol{k\cdot x}}\chi_{\tau}(\k,t)b^{\dag}_{\tau}(\k)\right],\label{eq:L1}
    \end{aligned}
\end{equation}
and
\begin{equation}
    \begin{aligned}
    &\dual{\lambda}(t,\x)=\int\frac{d^{3}k}{(2\pi)^{3}}\frac{1}{\sqrt{2mka^{3}(t)}}\sum^{4}_{\tau=1}
    \left[e^{-i\boldsymbol{k\cdot x}}\dual{\xi}_{\tau}(\k,t)a^{\dag}_{\tau}(\k)\right.\\
    &\left.+e^{i\boldsymbol{k\cdot x}}\dual{\chi}_{\tau}(\k,t)b_{\tau}(\k)\right].\label{eq:L2}
\end{aligned}
\end{equation}
To find exact solutions for Elko and its dual, it suffices to solve~(\ref{eq:eom}). After obtaining the exact solutions for $\xi_{\tau},\chi_{\tau}$, we can use the definitions~(\ref{eq:dual_xi1}-\ref{eq:dual_zetz4}) to obtain $\dual{\xi}_{\tau},\dual{\chi}_{\tau}$. 
% thus making their forms different from the Minkowski counterparts. 
%Substituting~(\ref{eq:curved_Elko}) into~(\ref{eq:Elko3}), we obtain
Substituting~(\ref{eq:L1}-\ref{eq:L2}) into~(\ref{eq:eom1}-\ref{eq:eom2}), we obtain 
\begin{equation}
    \begin{aligned}
&\ddot{\xi}_{\tau}(\k,t)+\left[\frac{k^{2}}{a^{2}}-\frac{3}{2}\left(\frac{\dot{a}^{2}}{a^{2}}+\frac{\ddot{a}}{a}\right)+i\frac{\dot{a}}{a^{2}}\gamma^{0}\gamma^{i}k_{i}+\right.\\
    &\left.\left(m^{2}+\frac{1}{4}R\right)\right]\xi_{\tau}(\k,t)=0,\label{eq:xi_eom}
\end{aligned}
\end{equation}   
and
\begin{equation}
    \begin{aligned}
    &\ddot{\chi}_{\tau}(\k,t)+\left[\frac{k^{2}}{a^{2}}-\frac{3}{2}\left(\frac{\dot{a}^{2}}{a^{2}}+\frac{\ddot{a}}{a}\right)-i\frac{\dot{a}}{a^{2}}\gamma^{0}\gamma^{i}k_{i}+\right.\\
    &\left.\left(m^{2}+\frac{1}{4}R\right)\right]\chi_{\tau}(\k,t)=0,\label{eq:chi_eom}
\end{aligned}
\end{equation}
where $k^{2}=\k\cdot\k$. Using the identity
\begin{equation}
    R=6\left(\frac{\ddot{a}}{a}+\frac{\dot{a}^{2}}{a^{2}}\right),
\end{equation}
the equations simplify to\footnote{It is instructive to note the following structures of the differential equations by writing them in terms of the Weyl spinors. Taking $\xi_{\tau}\equiv(\begin{matrix}\psi_{\tau} & \phi_{\tau}\end{matrix})$
where $\psi_{\tau}$ and $\phi_{\tau}$ are the right- and left-handed Weyl spinors respectively,~(\ref{eq:xi_eom}) becomes
\begin{equation}
    \ddot{\psi}_{\tau}(\k,t)+\left(\frac{k^{2}}{a^{2}}+m^{2}\right)\psi_{\tau}-i\frac{\dot{a}}{a^{2}}(\s\cdot\k)\psi_{\tau}(\k,t)=0,\label{eq:eom_psi} 
\end{equation}
and
\begin{equation}
    \ddot{\phi}_{\tau}(\k,t)+\left(\frac{k^{2}}{a^{2}}+m^{2}\right)\phi_{\tau}+i\frac{\dot{a}}{a^{2}}(\s\cdot\k)\phi_{\tau}(\k,t)=0.\label{eq:eom_phi}
\end{equation}

If we follow the traditional construct by taking $\psi_{\tau}=\vartheta\Theta\phi^{*}_{\tau}$, then~(\ref{eq:eom_psi}-\ref{eq:eom_phi}) would reduce to $(\s\cdot\k)\phi_{\tau}=0$ which only has trivial solution $\phi_{\tau}=0$. This cannot be correct because in the flat space limit, it is impossible for the trivial solution to approach the non-trivial solutions in Minkowski space. The resolution to this problem is simple -  the left- and right-handed components of Elko must be independent.
}

\begin{equation}
    \ddot{\xi}_{\tau}(\k,t)+\left[\frac{k^{2}}{a^{2}}+m^{2}+i\frac{\dot{a}}{a^{2}}\gamma^{0}\gamma^{i}k_{i}\right]\xi_{\tau}(\k,t)=0 \label{eq:xi_eom},
\end{equation}
    and
    \begin{equation}
    \ddot{\chi}_{\tau}(\k,t)+\left[\frac{k^{2}}{a^{2}}+m^{2}-i\frac{\dot{a}}{a^{2}}\gamma^{0}\gamma^{i}k_{i}\right]\chi_{\tau}(\k,t)=0.\label{eq:zeta_eom}
\end{equation}
Since $\xi_{\tau}$ and $\chi_{\tau}$ are four-component spinors, both~(\ref{eq:xi}-\ref{eq:zeta_eom}) are systems of coupled differential equations. Solving these equations directly is difficult. Fortunately, these equations can be decoupled. To do so, we impose the ansatz that Elko in curved space-time can be factorized into  a product of Elko in the Minkowski space-time and time-dependent scalar functions to be solved. As a result of the ansatz, $\xi_{1,4}(\k,t),\chi_{2,3}(\k,t)$ and $\xi_{2,3}(\k,t),\chi_{1,4}(\k,t)$ are eigenspinors of $\gamma^{0}\gamma^{i}k_{i}$ with eigenvalues $+k$ and $-k$ respectively. For Elko to have the correct time evolution, the time-dependent functions are chosen as follow
\begin{equation}
    \begin{aligned}
\xi_{1}(\k,t)&\equiv w_{1}(k,t)\xi_{1}(\k),&\quad \xi_{2}(\k,t)&\equiv w^{*}_{2}(k,t)\xi_{2}(\k),\\
\xi_{3}(\k,t)&\equiv w^{*}_{2}(k,t)\xi_{3}(\k),&\quad \xi_{4}(\k,t)&\equiv w_{1}(k,t)\xi_{4}(\k),\\
\end{aligned}
\end{equation}

and
\begin{equation}
    \begin{aligned}
\chi_{1}(\k,t)&\equiv w_{2}(k,t)\chi_{1}(\k),&\quad \chi_{2}(\k,t)&\equiv w^{*}_{1}(k,t)\chi_{2}(\k),\\
\chi_{3}(\k,t)&\equiv w^{*}_{1}(k,t)\chi_{3}(\k),&\quad \chi_{4}(\k,t)&\equiv w_{2}(k,t)\chi_{4}(\k),
\end{aligned}
\end{equation}
where $w_{1}$ and $w_{2}$ are scalar functions that satisfy
\begin{equation}
    \begin{aligned}
\ddot{w}_{1}(k,t)+\left[\frac{k^{2}}{a^{2}}+m^{2}+ik\frac{\dot{a}}{a^{2}}\right]w_{1}(k,t)&=0, \\
\ddot{w}^{*}_{2}(k,t)+\left[\frac{k^{2}}{a^{2}}+m^{2}-ik\frac{\dot{a}}{a^{2}}\right]w^{*}_{2}(k,t)&=0.
\end{aligned} 
\end{equation}

We now solve for $w_{1,2}$ in the dS space with $a(t)=e^{Ht}$ where $H$ is the Hubble constant. Making the following change of variables
\begin{equation}
z=2ik\eta,\quad \eta=-\frac{e^{-Ht}}{H},
\end{equation}
where $\eta$ is the conformal time and adopt the normalizations
\begin{equation}
w_{1}(k,t)\equiv z^{-1/2}\varphi(k,z),\quad 
w_{2}(k,t)\equiv z^{-1/2}\psi(k,z),
\end{equation}
we obtain
\begin{equation}
   \begin{aligned}
    \frac{d^{2}\varphi}{dz^{2}}+\left[\left(\frac{1}{4}+\frac{m^{2}}{H^{2}}\right)\frac{1}{z^{2}}-\frac{1}{4}-\frac{1}{2z}\right]\varphi=&0,\\
    \frac{d^{2}\psi^{*}}{dz^{2}}+\left[\left(\frac{1}{4}+\frac{m^{2}}{H^{2}}\right)\frac{1}{z^{2}}+\frac{1}{4}+\frac{1}{2z}\right]\psi^{*}=&0.\label{eq:psi}
\end{aligned} 
\end{equation}

The solutions to~(\ref{eq:psi}) are given by the Whittaker functions~\cite{NIST:DLMF}. In the flat space limit $\eta H\sim-1$, we require $w_{1}\sim e^{-ik\eta}$ and $w_{2}\sim e^{ik\eta}$. With these limits in mind, we choose the solutions to be\footnote{The solution for $\psi^{*}(k,z)$ is $W_{\frac{1}{2},\mu}(z)$. Using the identity $W^{*}_{\frac{1}{2},\mu}(z)=W_{\frac{1}{2},-\mu}(-z)$, we obtain $w_{2}$.}
\begin{equation}
\begin{aligned}
&w_{1}(k,t)=c(k)z^{-1/2}W_{-\frac{1}{2},-\mu}(z),\\
&w_{2}(k,t)=d(k)z^{-1/2}W_{\frac{1}{2},-\mu}(-z),
\end{aligned}
\end{equation}
where $\mu\equiv im/H$ and $c,d$ are constants to be determined. 
For the dual spinors, we adopt the same definitions in the Minkowski space-time~(\ref{eq:dual_xi1}-\ref{eq:dual_zetz4}) to obtain
\begin{equation}
    \begin{aligned}
\dual{\xi}_{1}(\k,t)&=w_{2}(k,t)\dual{\xi}_{1}(\k),\quad &
\dual{\xi}_{2}(\k,t)&=w^{*}_{1}(k,t)\dual{\xi}_{2}(\k),\\
\dual{\xi}_{3}(\k,t)&=w^{*}_{1}(k,t)\dual{\xi}_{3}(\k),\quad &
\dual{\xi}_{4}(\k,t)&=w_{2}(k,t)\dual{\xi}_{4}(\k),
\end{aligned}
\end{equation}

and
\begin{equation}
    \begin{aligned}
\dual{\chi}_{1}(\k,t)&=w_{1}(k,t)\dual{\chi}_{1}(\k),\quad &
\dual{\chi}_{2}(\k,t)&=w^{*}_{2}(k,t)\dual{\chi}_{2}(\k),\\
\dual{\chi}_{3}(\k,t)&=w^{*}_{2}(k,t)\dual{\chi}_{3}(\k),\quad &
\dual{\chi}_{4}(\k,t)&=w_{1}(k,t)\dual{\chi}_{4}(\k).
\end{aligned}
\end{equation}

The Elko solutions and their duals are chosen in such a way as to make the computations of the canonical anti-commutators more expedient. We now compute $\{\lambda(t,\x),\dual{\lambda}(t,\y)\}$ and $\left\{\lambda(t,\x),\pi(t,\y)\right\}$. The first one  evaluates to
\begin{equation}
\begin{aligned}
&\left\{\lambda(t,\x),\dual{\lambda}(t,\y)\right\}=(2\pi)^{-3}\int\frac{d^{3}k}{2mka^{3}(t)}\\
&\sum^{4}_{\tau=1}
\left[\xi_{\tau}(\k,t)\dual{\xi}_{\tau}(\k,t)+\chi_{\tau}(-\k,t)\dual{\chi}(-\k,t)\right].
\end{aligned}
\end{equation}
The relevant spin-sums are
\begin{equation}
\begin{aligned}
&\sum^{4}_{\tau=1}\xi(\k,t)\dual{\xi}(\k,t)=\left[\begin{matrix}
A(\k,t) & \mathbb{O}_{2} \\
\mathbb{O}_{2} & A^{\dag}(\k,t)
\end{matrix}\right],\\
&\sum^{4}_{\tau=1}\chi(\k,t)\dual{\chi}(\k,t)=\left[\begin{matrix}
B(\k,t) & \mathbb{O}_{2} \\
\mathbb{O}_{2} & B^{\dag}(\k,t)
\end{matrix}\right],\label{eq:A}
\end{aligned}
\end{equation}
where
\begin{equation}
    \begin{aligned}
&A(\k,t)=c(k)d(k)m\\
&\left[\begin{matrix}
w^{*}_{1}w^{*}_{2}c^{2}_{\theta/2}+w_{1}w_{2}s^{2}_{\theta/2} & e^{-i\phi}c_{\theta/2}s_{\theta/2}(w^{*}_{1}w^{*}_{2}-c.c.) \\
e^{i\phi}c_{\theta/2}s_{\theta/2}(w^{*}_{1}w^{*}_{2}-c.c.) & w^{*}_{1}w^{*}_{2}c^{2}_{\theta/2}+w_{1}w_{2}s^{2}_{\theta/2}
\end{matrix}\right],\\
&B(\k,t)=c(k)d(k)m\\
&\left[\begin{matrix}
-w_{1}w_{2}c^{2}_{\theta/2}-w^{*}_{1}w^{*}_{2}s^{2}_{\theta/2} & e^{-i\phi}c_{\theta/2}s_{\theta/2}(w^{*}_{1}w^{*}_{2}-c.c.) \\
e^{i\phi}c_{\theta/2}s_{\theta/2}(w^{*}_{1}w^{*}_{2}-c.c.) & -w^{*}_{1}w^{*}_{2}c^{2}_{\theta/2}-w_{1}w_{2}s^{2}_{\theta/2}
\end{matrix}\right],
\end{aligned}
\end{equation}

where $c_{\theta/2}=\cos(\theta/2),s_{\theta/2}=\sin(\theta/2)$. The two matrices are related by $A(\k,t)=-B(-\k,t)$ so we obtain
\begin{equation}
\left\{\lambda(t,\x),\dual{\lambda}(t,\y)\right\}=\mathbb{O}_{4}.\label{eq:L_pi}
\end{equation}
In obtaining~(\ref{eq:A}-\ref{eq:L_pi}), we chose $cd$ to be real, making it a global multiplicative factor in the spin-sums. If $cd$ is complex, then the anti-commutator~(\ref{eq:L_pi}) would be non-vanishing. We will determine $c$ and $d$ by demanding Elko to have the correct time evolution in the flat space limit. 

From the Lagrangian density, the conjugate momentum is
$\pi=a^{3}\partial_{t}\dual{\lambda}$ so we obtain
\begin{equation}
\begin{aligned}
&\left\{\lambda(t,\x),\pi(t,\y)\right\}=(2\pi)^{-3}\int\frac{d^{3}k}{2mk}\\
&\sum^{4}_{\tau=1}
\left[\xi_{\tau}(\k,t)\dot{\dualn{\xi}}_{\tau}(\k,t)+\chi_{\tau}(-\k,t)\dot{\dualn{\chi}}_{\tau}(-\k,t)\right].\label{eq:Lpi}
\end{aligned}
\end{equation}
The spin-sum in~(\ref{eq:Lpi}) is
\begin{equation}
\begin{aligned}
&\sum^{4}_{\tau=1}
\left[\xi_{\tau}(\k,t)\dot{\dualn{\xi}}_{\tau}(\k,t)+\chi_{\tau}(-\k,t)\dot{\dualn{\chi}}_{\tau}(-\k,t)\right]\\
&=
\left[\begin{matrix}
C(\k,t) & \mathbb{O}_{2} \\
\mathbb{O}_{2} & -C^{\dag}(\k,t)
\end{matrix}\right],
\end{aligned}
\end{equation}
where
\begin{equation}
 C(\k,t)=m\left[\begin{matrix}
 \mathcal{W}s^{2}_{\theta/2}-\mathcal{W}^{*}c^{2}_{\theta/2} & -e^{-i\phi}(\mathcal{W}+\mathcal{W}^{*}) \\
-e^{i\phi}(\mathcal{W}+\mathcal{W}^{*}) & c^{2}_{\theta/2}\mathcal{W}-s^{2}_{\theta/2}\mathcal{W}^{*}
 \end{matrix}\right]
\end{equation}
with $\mathcal{W}$ being the Wronskian of $w_{1}$ and $w_{2}$. Using result from~\cite{NIST:DLMF}, we obtain
\begin{align}
\mathcal{W}&\equiv w_{1}\dot{w}_{2}-\dot{w}_{1}w_{2}\nonumber\\
&=-c(k)d(k)H\mathcal{W}\left\{W_{-\frac{1}{2},-\mu}(z),W_{\frac{1}{2},-\mu}(-z)\right\}\nonumber\\
&=ic(k)d(k)H
\end{align}
and
\begin{equation}
\left\{\lambda(t, \x), \pi(t, \x')\right\}=\frac{i}{(2\pi)^{3}}\int\frac{d^{3}k}{2k}e^{i\boldsymbol{k\cdot(x-y)}}c(k)d(k)H\mathbb{I}_{4},
\end{equation}
To obtain the correct anti-commutator, we must have $cd=2k/H$ so that
\begin{equation}
\left\{\lambda(t, \x), \pi(t, \x')\right\}=i\delta^{3}(\x-\y)\mathbb{I}_{4}.
\end{equation}
The asymptotic limit of $w_{1}$ and $w^{*}_{2}$ in the flat space limit $\eta H\sim-1$ are
\begin{equation}
w_{1}(k,t)\sim -\frac{i}{2k\eta}c(k)e^{-ik\eta},\quad
w^{*}_{2}(k,t)\sim id^{*}(k)e^{-ik\eta}.
\end{equation}
Choosing
\begin{equation}
c(k)=-\frac{2ik}{H},\quad d(k)=i,
\end{equation}
we obtain the desired behaviours $w_{1}\sim e^{-ik\eta},w^{*}_{2}\sim e^{-ik\eta}$.

\section{Conclusion and outlook}\label{conc}

The foundation of Elko in Minkowski space-time was established in~\cite{Ahluwalia:2022yvk}. Building upon this work, we initiate a systematic study of Elko in curved space-time. 

In this work, we study Elko in the spatially flat FLRW space-time following~\cite{HoffdaSilva:2014tth}. By working in the helicity basis and utilizing the dual-helicity structure of the spinors, we found exact solutions in the dS background and showed that the resulting mass dimension one fermionic fields satisfy the canonical anti-commutation relations. 
%These results establish the foundation for Elko in the dS background allowing for future works in curved space-time.

%In this work, we have first revisited the free elko fermion in flat spacetime, then we discuss the free Elko propagating in curved spacetime. Then we discuss the quantization of Elko fermion in dS spacetime. 

There are many open questions to be addressed. For more realistic situations such as inflation, it is necessary to study Elko in the quasi dS background. There are two cosmological models worthy of future research: (1) A model of scalar inflaton coupled to Elko. (2) Elko as an inflaton. The latter possibility was initially considered by Boehmer et al~\cite{Boehmer:2006qq,Boehmer:2007dh,Boehmer:2008ah,Boehmer:2008rz,Boehmer:2009aw,Boehmer:2010ma} though at that time, the correct degrees of freedom have not yet been identified and these works mainly focused on the classical dynamics. Since Elko is a dark matter candidate, the second proposal is a model of dark matter driven inflation. In this case, it may be possible to connect inflation with the creation of dark matter.

In both scenarios, interactions coming from the scalar, vector and tensor perturbations of the space-time metric can be derived using the ADM formalism~\cite{Arnowitt:1962hi}. The real challenge lies in computing observables such as multi-point correlation functions and particle creation rates. As noted in~\cite{Ahluwalia:2023slc}, the Elko interactions in the Minkowski space-time are in general pseudo Hermitian. This feature is also present in curved space-time. To compute observables, a formalism to deal with pseudo Hermitian operators is required. In the Minkowski space-time, the pseudo Hermiticity structure of the interacting Hamiltonian allows one to define new inner-product that preserves time translation symmetry and generalized unitarity though further works are required to ascertain whether observables can be consistently computed. Formalism and techniques developed in flat space can in principle be generalized to curved space though subtleties may arise since time translation is no longer a symmetry. 
%Here, it may be instructive to first consider the Schwinger-Keldysh formalism with pseudo Hermitian Hamiltonians in flat space before attempting its generalizations to curved space when studying non-linear perturbations of the space-time metric.

Another open question is how Elko affects torsion in space-time. In Einstein-Cartan gravity, which is the gauge theory of the Poincar\'{e} group~\cite{Kibble:1961ba,Blagojevic:2013xpa}, the spin density of matter can cause torsion~\cite{Hehl:1976kj,Medina:2018rnl}. Since the Dirac and Elko field have different spin density tensors,  it can be expected that the two fields will induce different torsion effects.

%Pseudo Hermitian gravitational interactions naturally arise from inflation, coming from the scalar and tensor perturbations of the space-time metric. 

%Secondly, possible cosmological signatures of Elko is worth investigating. This includes non-Gaussianities and primordial gravitational waves.

%Thirdly, the possibility that the elko fermion itself can be a candidate of inflaton is very interesting, especially if they can be distinguished from the single field inflation observationally.

%Last but not the least, since elko is a dark matter candidate, we would like to study the percentage of the dark matter it can account for. So we would like to study the particle creation rate of this type of dark matter and see if it can account for the whole of the dark matter relic abundance. 

\appendix 

\section{The Lichnerowicz formula}~\label{Lichnerowicz}

We derive the Lichnerowicz formula following~\cite{Shapiro:2016pfm}. Since $\Gamma_\mu$ satisfy $[\Gamma_\mu,\gamma^\nu]=\nabla_\mu\gamma^\nu $, which means $\nabla_\mu$ commute with $\gamma^\nu $, we write $\nabla_{\mu}\nabla_{\nu}$ as a sum of its anti-commutator and commutator
\begin{align}
    \gamma^{\mu}\nabla_{\mu}\gamma^{\nu}\nabla_{\nu}\lambda
    &=\frac{1}{2}\gamma^{\mu}\gamma^{\nu}\left\{\nabla_{\mu},\nabla_{\nu}\right\}\lambda+\frac{1}{2}\gamma^{\mu}\gamma^{\nu}\left[\nabla_{\mu},\nabla_{\nu}\right]\lambda\nonumber\\
    &=g^{\mu\nu}\nabla_{\mu}\nabla_{\nu}\lambda+\frac{1}{2}\gamma^{\mu}\gamma^{\nu}\left[\nabla_{\mu},\nabla_{\nu}\right]\lambda.\label{eq:Laplacian}
\end{align}
Our task is to evaluate the commutator on the third line of~(\ref{eq:Laplacian}). Because $\nabla_{\nu}\lambda$ is a spin-vector, its covariant derivative contains both the spin connection and the Christoffel symbol
\begin{equation}
\begin{aligned}
\nabla_\mu \nabla_\nu \lambda= & \left(\partial_\mu \Gamma_\nu\right) \lambda+\Gamma_\nu \partial_\mu \lambda+\partial_\mu \partial_{\nu \lambda}+\Gamma_\mu \partial_\nu \lambda+\Gamma_\mu \Gamma_\nu \lambda \\
+ & \Gamma_{\mu \nu}^p \partial_\chi \lambda+\Gamma_{\mu \nu} \Gamma_\chi \lambda
\end{aligned}
\end{equation}
the commutator reads
\begin{equation}
\left[\nabla_\mu, \nabla_\nu\right] \lambda=\left(\partial_\mu \Gamma_\nu-\partial_\nu \Gamma_\mu+\Gamma_\mu \Gamma_\nu-\Gamma_\nu \Gamma_\mu\right) \lambda
\end{equation}
substitute the definition of $\Gamma_\mu$, we get
\begin{equation}
\begin{aligned}
&\gamma^\mu \gamma^\nu\left[\nabla_\mu, \nabla_\nu\right] \lambda=\left\{\frac{1}{4}\left(\partial_ \mu {\omega_\nu}^{a b}-\partial_\nu {\omega_\mu}^{a b}\right) \gamma^\mu \gamma^\nu \gamma_a \gamma_b+\right.\\
&\left.\frac{1}{16} {{\omega_\mu}^{a}}_c {\omega_\nu}^{c b} \gamma^\mu \gamma^\nu\left[\gamma_a \gamma_b, \gamma_c \gamma_d\right]\right\} \lambda
\end{aligned}
\end{equation}
Using the identity
\begin{equation}
    [\gamma_a \gamma_b, \gamma_c \gamma_d]=
    2(\eta_{b c}\gamma_{a}\gamma_{d}-\eta_{a c}\gamma_{b}\gamma_{d}+\eta_{b d}\gamma_{c}\gamma_{a}-\eta_{a d}\gamma_{c}\gamma_{b}),
\end{equation}
we obtain
\begin{equation}
\begin{aligned}
&\gamma^\mu \gamma^\nu\left[\nabla_\mu, \nabla_\nu\right] \lambda \\
&=\frac{1}{2} e_{a \chi} e_{b \sigma}\left(\partial _{[\mu} \omega_{\nu]}^{a b}+{{\omega_{[\mu} }^a} _c \omega_{\nu]}{ }^{c b}\right) \gamma^\mu \gamma^\nu \gamma^\chi \gamma^\sigma \lambda,
\end{aligned}
\end{equation}
use equation (74) of \cite{yepez2011einsteins}
\begin{equation}
    \gamma^\mu \gamma^\nu\left[\nabla_\mu, \nabla_\nu\right] \lambda =\frac{1}{4}R_{\mu\nu\chi\sigma}\gamma^{\mu}\gamma^{\nu}\gamma^{\chi}\gamma^{\sigma}\lambda.
\end{equation}
Equation~(\ref{eq:Laplacian}) becomes
\begin{equation}
    \gamma^{\mu}\nabla_{\mu}\gamma^{\nu}\nabla_{\nu}\lambda=g^{\mu\nu}\nabla_{\mu}\nabla_{\nu}\lambda+\frac{1}{8}R_{\mu\nu\chi\sigma}\gamma^{\mu}\gamma^{\nu}\gamma^{\chi}\gamma^{\sigma}\lambda. \label{eq:Laplacian2}
\end{equation}
The second term of~(\ref{eq:Laplacian2}) can be simplified by considering
\begin{align}
    R_{\mu\nu\chi\sigma}\gamma^{\nu}\gamma^{\chi}\gamma^{\sigma}&=-2R_{\mu\nu}\gamma^{\nu}-R_{\mu\nu\chi\sigma}\gamma^{\chi}\gamma^{\nu}\gamma^{\sigma}\nonumber\\
    &=-2R_{\mu\nu}\gamma^{\nu}+(R_{\mu\sigma\nu\chi}+R_{\mu\chi\sigma\nu})\gamma^{\chi}\gamma^{\nu}\gamma^{\sigma}
\end{align}
where $R_{\mu\nu}={R^{\sigma}}_{\mu\sigma\nu}$ and we have used the Bianchi identity. Using the identity
\begin{equation}
    \gamma^{\chi}\gamma^{\nu}\gamma^{\sigma}=\gamma^{\sigma}\gamma^{\chi}\gamma^{\nu}-2g^{\chi\sigma}\gamma^{\nu}+2g^{\nu\sigma}\gamma^{\chi}
\end{equation}
we obtain
\begin{equation}
    R_{\mu\nu\chi\sigma}\gamma^{\nu}\gamma^{\chi}\gamma^{\sigma}=-2R_{\mu\nu}\gamma^{\nu}
\end{equation}
and hence
\begin{align}
    R&=g^{\mu\nu}R_{\mu\nu}\nonumber\\
    &=-\frac{1}{2}R_{\mu\nu\chi\sigma}\gamma^{\mu}\gamma^{\nu}\gamma^{\chi}\gamma^{\sigma}.\label{eq:R}
\end{align}
Substituting~(\ref{eq:R}) into~(\ref{eq:Laplacian2}), we obtain the Lichnerowicz formula
\begin{equation}
    \gamma^{\mu}\nabla_{\mu}\gamma^{\nu}\nabla_{\nu}\lambda=\left(g^{\mu\nu}\nabla_{\mu}\nabla_{\nu}-\frac{1}{4}R\right)\lambda.
\end{equation}

\acknowledgments

We would like to thank Chon-Man Sou for discussions in the early stage of this work. We also thank Yanjiao Ma and Yu Zhang for helpful discussions. SZ is supported by Natural Science Foundation of China under Grant No.12147102 at Chongqing University.  

\bibliography{apssamp}% Produces the bibliography via BibTeX.

\end{document}